\documentstyle[12pt]{article}
\topmargin - 1.5 cm 

\begin{document}

\begin{center}
{\large{\bf 8D OSCILLATOR AS A HIDDEN SU(2) -- MONOPOLE}}
\end{center}
\begin{center}
{\bf L.G. Mardoyan${}^1$, A.N. Sissakian${}^2$,
V.M. Ter--Antonyan${}^3$ }
\end{center}
\begin{center}
{\large{\bf Joint Institute for Nuclear Research}}
\end{center}
\begin{center}
{\bf 141980, Dubna, Moscow Region, Russia}
\end{center}

\footnotetext[1]{E-mail: mardoyan@thsun1.jinr.dubna.su}
\footnotetext[2]{E-mail: sisakian@jinr.dubna.su}
\footnotetext[3]{E-mail: terant@thsun1.jinr.dubna.su}

\begin{center}
\underline{Abstract}
\end{center}
{\small In this report, in the framework of an analytical approach
and with help of the generalized version of the Hurwitz transformation
the five--dimensional $SU(2)$--monopole model is constructed from the
eight--dimensional quantum oscillator. The $SU(2)$--monopole fields,
the Clebsh--Gordan expansion stimulated by the space--gauge coupling,
the hyperangle and the radial parts of the total wave function, the
energy spectrum of the charge--monopole bound system and the corresponding
degeneracy are calculated.}

\vspace{1cm}
{\large{\bf 1. Introduction.}}
\vspace{0.3cm}

This paper deals with the problem of monopole generation from
oscillator--like systems, i.e. systems with a potential chosen as
"oscillator + anything". In turn, the mentioned problem is connected with
the search for the electromagnetic duality (ED) in the structure of
Quantum Mechanics (QM). The existence of QM--duality seems important
for two reasons. First, QM is a mathematically more simple theory than
the gauge theories, so we have an excellent polygon for experience in
ED. Second, there appears a wide range of applications because of ED
pretentions to realize accurate calculations outside
perturbation theory: according to ED, strongly coupled gauge theories
can be formulated in the form of weakly coupled magnetic monopoles [1].

During the last years, the following machinery has been developed for
a monopole generation: Hurwitz--like transformations applied to 2D, 4D
and 8D quantum oscillators transfer them into the charge--monopole bound
systems in $R^2$, $R^3$ and $R^5$, respectively [2-4]. In two space
dimensions the oscillator model was also constructed which can be
transformed into a charge--monopole bound system with fractional
statistics, interpolating the bosonic and fermionic limits [5].
Thus, the important extension of ED to the world of anyons is achieved.

Recently, the algebraic approach has been developed to clarify the
relation between the 8D quantum oscillator and the charge--dyon bound
system with the $SU(2)$--monopole [6]${}^4$. This approach is exhuastive
\footnotetext[4]{The particular version of the problem was previously
considered in [7].}
but rather abstract. We make here
an attempt to fulfill this gap${}^5$ by presenting the analytical
\footnotetext[5]{See also [2] where the analytical approach to the same
problem was presented more concisely.}
approach that is more explicit and hence more acceptable for understanding.
Special attention is given to the $SU(2)$--monopole fields, the
space--gauge coupling and to the spectroscopy of the charge--dyon
bound system.

\vspace{0.5cm}
{\large{\bf 2. U(1)--Monopole.}}
\vspace{0.3cm}

Let us recall the way used for passage from the 4D oscillator to the
3D $U(1)$--monopole. The initial system is governed by the equation
\begin{equation}
\frac{{\partial}^2\psi}{\partial u^2_\mu}
+ \frac{2M}{\hbar^2}\left(E-\frac{M\omega^2u^2}{2}\right)\psi=0,
\end{equation}
where $u_{\mu}\in R^4$, $\mu=0,1,2,3$; $u^2=u_{\mu}u_{\mu}$.

With the help of the special transformation${}^6$
\footnotetext[6]{Notice that the three lines of (2) are the
KS--transformation well known from the Celestial Mechanics [8].}
\begin{eqnarray}
x_1 &=& 2(u_0u_2 + u_1u_3) \nonumber \\ [3mm]
x_2 &=& 2(u_0u_3 - u_1u_2) \nonumber \\ [3mm]
x_3 &=& u_0^2 + u_1^2 - u_2^2 - u_3^2 \\ [3mm]
\gamma &=& i\ln{\frac{u_0-iu_1}{u_0+iu_1}} \nonumber
\end{eqnarray}
we present $R^4$ as a direct product $R^3\otimes S_{4\pi}^1$ of the
new configuration space $R^3$ with the Cartesian coordinates
$x_j \in (-\infty, \infty)$ and the intrinsic space $S_{4\pi}^1$ with
the coordinate $\gamma \in [0,4\pi)$. In the new coordinates, Eq.(1) can
be led to the form
\begin{eqnarray*}
\frac{1}{2M}\left(-i\hbar \frac{\partial}{\partial x_j}-
\hbar A_j \hat S\right)^2\psi + \frac{\hbar^2}{2Mr^2}
{\hat S}^2\psi - \frac{e^2}{r}\psi = \epsilon \psi
\end{eqnarray*}
Here $r=\sqrt{x_1^2+x_2^2+x_3^2}$, $\hat S = -i\partial/{\partial \gamma}$,
\begin{eqnarray}
\epsilon = -M\omega^2 /8,\,\,\,\,\,\,\,\,\, e^2 = E/4,
\end{eqnarray}
and
\begin{eqnarray*}
\vec A = \frac{1}{r(r + x_3)}(-x_2, x_1, 0)
\end{eqnarray*}
As we can to see, Eq.(1) describes the charge--dyon system with the
$U(1)$--monopole${}^7$.
\footnotetext[7]{The successive theory for the charge--dyon system
with the $U(1)$--monopole and its dynamic symmetry was initially
constructed by Zwanziger [9].}

\vspace{0.5cm}
{\large{\bf 3. SU(2)--Monopole.}}
\vspace{0.3cm}

The described scheme can be generalized to the 8D isotropic
quantum oscillator in the following way. Let $u_{\mu}$ is the
Cartesian coordinates of $R^8$, $u^2=u_{\mu}u_{\mu}$ and
$\mu = 0,1,...7$.
\begin{equation}
\frac{{\partial}^2\psi}{\partial u^2_\mu}
+ \frac{2M}{\hbar^2}\left(E-\frac{M\omega^2u^2}{2}\right)\psi=0,
\,\,\,\,\,\,\,u_{\mu}\in R^8
\end{equation}
Consider the transformation${}^8$
\footnotetext[8]{Formulae (5) is known as the Hurwitz transformation [10];
(6) is copied from [11].}
\begin{eqnarray}
x_0 &=& u_0^2 + u_1^2 + u_2^2 + u_3^2
-u_4^2 - u_5^2- u_6^2 - u_7^2 \nonumber \\  [2mm]
x_1 &=& 2(u_0u_4 + u_1u_5 - u_2u_6 - u_3u_7) \nonumber \\  [2mm]
x_2 &=& 2(u_0u_5 - u_6u_4 + u_2u_7 - u_3u_6)  \\  [2mm]
x_3 &=& 2(u_0u_6 + u_1u_7 + u_2u_4 + u_3u_5) \nonumber \\  [2mm]
x_4 &=& 2(u_0u_7 - u_1u_6 - u_2u_5 + u_3u_4) \nonumber
\end{eqnarray}
\begin{eqnarray}
{\alpha}_T &=& \frac{i}{2}\ln{\frac{(u_0 + iu_1)(u_2 - iu_3)}
{(u_0 - iu_1)(u_2 + iu_3)}}\in[0,2\pi) \nonumber \\ [2mm]
{\beta}_T &=& 2\arctan{\left(\frac{u_2^2 + u_3^2}
{u_0^2 + u_1^2}\right)^{1/2}}\in[0,\pi] \\ [2mm]
{\gamma}_T &=& \frac{i}{2}\ln{\frac{(u_0 + iu_1)(u_2 + iu_3)}
{(u_0 - iu_1)(u_2 - iu_3)}}\in[0,4\pi) \nonumber
\end{eqnarray}
It follows from (5) and (6) that
\begin{eqnarray*}
\Delta_8 = 4r\left(\Delta_5
- 2iA^a_j{\hat T}_a
\frac{\partial}{\partial x_j}
- \frac{2}{r(r + x_0)}{\hat T}^2 \right)
\end{eqnarray*}
Here $r=(x_jx_j)^{1/2}$, $\Delta_8={\partial}^2/{\partial u_{\mu}^2}$,
$\Delta_5={\partial}^2/{\partial x_j^2}$,
\begin{eqnarray}
\hat T_1 &=& i\left(\cos{{\alpha}_T}\cot{{\beta}_T}
\frac{\partial}{\partial {\alpha}_T}+
\sin{{\alpha}_T}\frac{\partial}{\partial {\beta}_T} -
\frac{\cos{{\alpha}_T}}{\sin{{\beta}_T}}
\frac{\partial}{\partial {\gamma}_T}\right) \nonumber \\ [3mm]
\hat T_2 &=& i\left(\sin{{\alpha}_T}\cot{{\beta}_T}
\frac{\partial}{\partial {\alpha}_T} -
\cos{{\alpha}_T}\frac{\partial}{\partial {\beta}_T} -
\frac{\sin{{\alpha}_T}}{\sin{{\beta}_T}}
\frac{\partial}{\partial {\gamma}_T}\right)  \\ [3mm]
\hat T_3 &=& -i\frac{\partial}{\partial {\alpha}_T} \nonumber
\end{eqnarray}
and
\begin{eqnarray*}
\vec A^1 &=& \frac{1}{r(r + x_0)}(0, -x_4, -x_3, x_2, x_1)  \\ [2mm]
\vec A^2 &=& \frac{1}{r(r + x_0)}(0, x_3, -x_4, -x_1, x_2)  \\ [2mm]
\vec A^3 &=& \frac{1}{r(r + x_0)}(0, x_2, -x_1, x_4, -x_3)
\end{eqnarray*}

Every term of the triplet $A_j^a$ coincides with the vector potential of 5D
Dirac monopole${}^9$
with a unit topological charge and the line of singularity along the
nonpositive part of the $x_0$--axis. The vectors $A_j^a$ are
\footnotetext[9]{The SU(2)--monopole theory in $R^5$ was constructed
by Yang [12].}
orthogonal to each other,
\begin{eqnarray*}
A^a_jA^b_j = \frac{1}{r^2}\frac{(r - x_0)}{(r + x_0)}
{\delta}_{ab}
\end{eqnarray*}
and also to the vector $\vec x=(x_0,x_1,x_2,x_3,x_4)$.

Returning to Eq.(4) we obtain
\begin{equation}
\left(\Delta_5
- 2iA^a_j{\hat T}_a
\frac{\partial}{\partial x_j}
- \frac{2}{r(r + x_0)}{\hat T}^2 \right)\psi +
\frac{2M}{\hbar^2}\left(\epsilon + \frac{e^2}{r}\right)\psi=0
\end{equation}
or in a more informative form
\begin{equation}
\frac{1}{2M}\left(-i\hbar \frac{\partial}{\partial x_j}-
\hbar A^a_j \hat T_a\right)^2\psi + \frac{\hbar^2}{2Mr^2}
{\hat T}^2\psi - \frac{e^2}{r}\psi = \epsilon \psi
\end{equation}

We see that Eq.(9) describes the charge--dyon bound system with the
$SU(2)$--monopole.

\vspace{1cm}
{\large{\bf 4. The Field Tensor.}}
\vspace{0.3cm}

Consider now the field tensor
\begin{eqnarray*}
F_{ik}^a ={\partial}_iA_k^a-{\partial}_kA_i^a+
i{\epsilon}_{abc}A_i^bA_k^c
\end{eqnarray*}
Since $A_0^a\equiv 0$, we have $(\mu, \nu = 1,2,3,4)$
\begin{eqnarray*}
A_{\mu}^a = \frac{2i}{r(r+x_0)}{\tau}_{\mu \nu}^a x_{\nu}
\end{eqnarray*}
where
\begin{eqnarray*}
{\tau}^1 = \frac{1}{2}\left(\begin{array}{cc}
0&i{\sigma}^1 \\
-i{\sigma}^1&0
\end{array} \right),\,\,\,
{\tau}^2 = \frac{1}{2}\left(\begin{array}{cc}
0&-i{\sigma}^3 \\
i{\sigma}^3&0
\end{array} \right),\,\,\,
{\tau}^3 = \frac{1}{2}\left(\begin{array}{cc}
i{\sigma}^2&0 \\
0&i{\sigma}^2
\end{array} \right)
\end{eqnarray*}
and ${\sigma}^a$ are the Pauli matrices
\begin{eqnarray*}
{\sigma}^1 = \left(\begin{array}{cc}
0 &1 \\
1 &0
\end{array} \right),\,\,\,
{\sigma}^2 = \left(\begin{array}{cc}
0 &-i \\
i &0
\end{array} \right),\,\,\,
{\sigma}^3 = \left(\begin{array}{cc}
1&0 \\
0&-1
\end{array} \right)
\end{eqnarray*}
With the help of these formulae we find that
\begin{eqnarray}
F_{0 \mu}^a = -\frac{2i}{r^3}{\tau}_{\mu \nu}^ax_{\nu} =
-\frac{r+x_0}{r^2}A_{\mu}^a
\end{eqnarray}
\begin{eqnarray*}
F_{\mu \nu}^a = \frac{2r+x_0}{r^2(r+x_0)}\left(x_{\nu}A_{\mu}^a
-x_{\mu}A_{\nu}^a\right) -
\frac{4i}{r(r+x_0)}{\tau}_{\mu \nu}^a +
{\epsilon}_{abc}A_{\mu}^bA_{\nu}^c
\end{eqnarray*}
It is a easy to verify that
\begin{eqnarray*}
[{\tau}^a,{\tau}^b] = i{\epsilon}_{abc}{\tau}^c
\end{eqnarray*}
\begin{eqnarray}
{\tau}_{\mu \lambda}^a{\tau}_{\lambda \nu}^b =
\frac{1}{4}{\delta}_{ab}{\delta}_{\mu \nu} +
\frac{i}{2}{\epsilon}_{abc}{\tau}_{\mu \nu}^c
\end{eqnarray}
Since ${\epsilon}_{abc}{\tau}_{\alpha \beta}^b{\tau}_{\mu \nu}^c$
is a pseudovector in $S^3$ and a tensor in $R^4$, we can construct
the general expansion
\begin{eqnarray*}
{\epsilon}_{abc}{\tau}_{\alpha \beta}^b{\tau}_{\mu \nu}^c =
A{\delta}_{\alpha \mu}{\tau}_{\nu \beta}^a+
B{\delta}_{\alpha \nu}{\tau}_{\mu \beta}^a+
C{\delta}_{\beta \nu}{\tau}_{\mu \alpha}^a+
D{\delta}_{\beta \mu}{\tau}_{\nu \alpha}^a+
E{\delta}_{\alpha \beta}{\tau}_{\mu \nu}^a+
F{\delta}_{\mu \nu}{\tau}_{\alpha \beta}^a
\end{eqnarray*}
Here, the l.h.s. is simmetric under indices
$\alpha$, $\beta$ and $\mu$, $\nu$, respectively. Hence,
$E=F=0$, $D=-A$, $C=-B$ and
\begin{eqnarray*}
{\epsilon}_{abc}{\tau}_{\alpha \beta}^b{\tau}_{\mu \nu}^c =
A\left({\delta}_{\alpha \mu}{\tau}_{\nu \beta}^a-
{\delta}_{\beta \mu}{\tau}_{\nu \alpha}^a\right) +
B\left({\delta}_{\alpha \nu}{\tau}_{\mu \beta}^a-
{\delta}_{\beta \nu}{\tau}_{\mu \alpha}^a\right)
\end{eqnarray*}
Further, $B=-A$, because after $\alpha \leftrightarrow \mu$
and $\beta \leftrightarrow \nu$ the l.h.s. changes the sign
but the r.h.s. changes its brackets
\begin{eqnarray*}
{\epsilon}_{abc}{\tau}_{\alpha \beta}^b{\tau}_{\mu \nu}^c =
A\left({\delta}_{\alpha \mu}{\tau}_{\nu \beta}^a-
{\delta}_{\alpha \nu}{\tau}_{\mu \beta}^a-
{\delta}_{\beta \mu}{\tau}_{\mu \alpha}^a+
{\delta}_{\beta \nu}{\tau}_{\mu \alpha}^a\right)
\end{eqnarray*}
Summing under $\beta$ and $\mu$ and taking into account Eq.(11),
we conclude that $A=i/2$; and therefore
\begin{eqnarray*}
{\epsilon}_{abc}{\tau}_{\alpha \beta}^b{\tau}_{\mu \nu}^c =
\frac{i}{2}\left({\delta}_{\alpha \mu}{\tau}_{\nu \beta}^a-
{\delta}_{\alpha \nu}{\tau}_{\mu \beta}^a+
{\delta}_{\beta \nu}{\tau}_{\mu \alpha}^a-
{\delta}_{\beta \mu}{\tau}_{\nu \alpha}^a\right)
\end{eqnarray*}
Using this formula one can easily show that
\begin{eqnarray*}
{\epsilon}_{abc}A_{\alpha}^b A_{\mu}^c =
\frac{1}{r(r+x_0)}\left(x_{\alpha}A_{\mu}^a-
x_{\mu}A_{\alpha}^a+2i\frac{r-x_0}{r}{\tau}_{\alpha \mu}^a\right)
\end{eqnarray*}
which leads to
\begin{eqnarray}
F_{\mu \nu}^a = \frac{1}{r^2}\left(x_{\nu}A_{\mu}^a -
x_{\mu}A_{\nu}^a - 2i{\tau}_{\mu \nu}^a\right)
\end{eqnarray}

It follows from (10), (12) and the explicit form of $A_j^a$ that
\begin{eqnarray*}
x_iF_{ij}^a \equiv 0
\end{eqnarray*}
Thus, both $A_j^a$ and $F_{jk}^a$ are orthogonal to the
vector $x_j$.

\vspace{1cm}
{\large{\bf 5. LT--Coupling.}}
\vspace{0.3cm}

Let us note that
\begin{eqnarray*}
iA_j^a\frac{\partial}{\partial x_j}=\frac{2}{r(r+x_0)}{\hat L}_a,
\end{eqnarray*}
where
\begin{eqnarray*}
\hat L_1 &=& \frac{i}{2}\left[D_{41}(x)+D_{32}(x)\right]  \\ [2mm]
\hat L_2 &=& \frac{i}{2}\left[D_{13}(x)+D_{42}(x)\right]  \\  [2mm]
\hat L_3 &=& \frac{i}{2}\left[D_{12}(x)+D_{34}(x)\right]
\end{eqnarray*}
and
\begin{eqnarray*}
D_{ij}(x) = - x_i\frac{\partial}{\partial x_j} +
x_j\frac{\partial}{\partial x_i}
\end{eqnarray*}
Using these formulae we can transform Eq.(8) into
\begin{equation}
\left(\Delta_5
- \frac{4}{r(r+x_0)}\hat {\vec L} \hat {\vec T}
- \frac{2}{r(r + x_0)}{\hat T}^2 \right)\psi +
\frac{2M}{\hbar^2}\left(\epsilon + \frac{e^2}{r}\right)\psi=0
\end{equation}
We see that Eq.(13) contains the {\it $LT$--coupling} term
demonstrating that we have no way to separate
the wave function dependence on $R^5$ and $S^3$.

Let us introduce in $R^5$ the hyperspherical coordinates $r\in [0,\infty)$,
$\theta \in [0,\pi]$, $\alpha \in [0,2\pi)$, $\beta \in [0,\pi]$,
$\gamma \in [0,4\pi)$ according to
\begin{eqnarray*}
x_0 &=& r\cos \theta  \nonumber \\ [2mm]
x_1 + ix_2 &=& r \sin \theta \cos \frac{\beta}{2}e^{i\frac{\alpha
+\gamma}{2}}  \\ [2mm]
x_3 + ix_4 &=& r \sin \theta \sin \frac{\beta}{2}e^{i\frac{\alpha -
\gamma}{2}}  \nonumber
\end{eqnarray*}
In these coordinates
\begin{eqnarray*}
\Delta_5 = \frac{1}{r^4}\frac{\partial}{\partial r}
\left(r^4 \frac{\partial}{\partial r}\right) +
\frac{1}{r^2 \sin^3 \theta}\frac{\partial}{\partial \theta}
\left(\sin^3 \theta \frac{\partial}{\partial \theta}\right) -
\frac{4}{r^2 \sin^2 \theta} {\hat L}^2
\end{eqnarray*}
where
\begin{eqnarray*}
{\hat L}_1 &=& i\left(\cos{\alpha}\cot{\beta}
\frac{\partial}{\partial \alpha}+
\sin{\alpha}\frac{\partial}{\partial \beta} -
\frac{\cos{\alpha}}{\sin{\beta}}
\frac{\partial}{\partial \gamma}\right) \nonumber \\ [3mm]
{\hat L}_2 &=& i\left(\sin{\alpha}\cot{\beta}
\frac{\partial}{\partial \alpha} -
\cos{\alpha}\frac{\partial}{\partial \beta} -
\frac{\sin{\alpha}}{\sin{\beta}}
\frac{\partial}{\partial \gamma}\right)  \\ [3mm]
{\hat L}_3 &=& -i\frac{\partial}{\partial \alpha} \nonumber
\end{eqnarray*}
and
\begin{eqnarray*}
{\hat L}^2 =
-\left[\frac{\partial^2}{\partial \beta^2}+
\cot \beta\frac{\partial}{\partial \beta}+
\frac{1}{\sin^2\beta}\left(\frac{\partial^2}{\partial {\alpha}^2}-
2\cos \beta \frac{\partial^2}{\partial \alpha \partial \gamma}+
\frac{\partial^2}{\partial {\gamma}^2}\right)\right]
\end{eqnarray*}
Let us introduce $\hat J_a = \hat L_a + \hat T_a$. Since
${\hat J}^2={\hat L}^2+{\hat T}^2+2{\hat L}_a{\hat T}_a$, Eq.(13)
can be rewritten as
\begin{equation}
\left(\Delta_{r \theta}
- \frac{{\hat L}^2}{r^2 \sin^2 \theta/2}
- \frac{{\hat J}^2}{r^2 \cos^2 \theta/2}\right)\psi +
\frac{2M}{\hbar^2}\left(\epsilon + \frac{e^2}{r}\right)\psi=0
\end{equation}
where
\begin{eqnarray*}
\Delta_{r \theta} =
\frac{1}{r^4}\frac{\partial}{\partial r}
\left(r^4 \frac{\partial}{\partial r}\right) +
\frac{1}{r^2 \sin^3 \theta}\frac{\partial}{\partial \theta}
\left(\sin^3 \theta \frac{\partial}{\partial \theta}\right)
\end{eqnarray*}

Emphasize that
\begin{eqnarray*}
[\hat L_a,\hat L_b] = i\epsilon_{abc}\hat L_c,\,\,\,\,\,\,
[\hat T_a,\hat T_b] = i\epsilon_{abc}\hat T_c,\,\,\,\,\,\,
[\hat J_a,\hat J_b] = i\epsilon_{abc}\hat J_c
\end{eqnarray*}

Introduce the separation ansatz
\begin{eqnarray*}
\psi=\Phi(r,\theta)G(\alpha,\beta,\gamma;\alpha_T,\beta_T,\gamma_T)
\end{eqnarray*}
where $G$ are the eigenfunctions of ${\hat L}^2$, ${\hat T}^2$ and
${\hat J}^2$ with the eigenvalues $L(L+1)$, $T(T+1)$ and $J(J+1)$. If
this is substituted into Eq.(14), the differential equation for
the function $\Phi(r,\theta)$ immediately follows
\begin{equation}
\left(\Delta_{r \theta}
- \frac{L(L+1)}{r^2 \sin^2 \theta/2}
- \frac{J(J+1)}{r^2 \cos^2 \theta/2}\right)\Phi +
\frac{2M}{\hbar^2}\left(\epsilon + \frac{e^2}{r}\right)\Phi=0
\end{equation}
Because of an $LT$--interaction, we look for the function $G$ in
the form
\begin{eqnarray*}
G =
\sum_{M=m+t}\left(JM|L,m';T,t'\right)D_{mm'}^L(\alpha,\beta,\gamma)
D_{tt'}^T(\alpha_T,\beta_T,\gamma_T)
\end{eqnarray*}
where $\left(JM|L,m';T,t'\right)$ are the Clebsh--Gordan coefficients
and $D_{mm'}^L$ and $D_{tt'}^T$ are the Wigner functions.

\vspace{1cm}
{\large{\bf 6. Hypermomentum.}}
\vspace{0.3cm}

Pick up the function $\Phi(r,\theta)$ of the form
\begin{eqnarray*}
\Phi(r,\theta) = R(r)Z(\theta)
\end{eqnarray*}
Then, equation (15) is separated into
\begin{eqnarray}
\frac{1}{\sin^3\theta}\frac{d}{d\theta}\left(\sin^3\theta
\frac{dZ}{d\theta}\right)-\frac{2L(L+1)}{1-\cos\theta}Z-
\frac{2J(J+1)}{1+\cos\theta}Z+\lambda(\lambda+3)Z=0
\end{eqnarray}
and a purely radial equation
\begin{eqnarray}
\frac{1}{r^4}\frac{d}{dr}\left(r^4\frac{dR}{dr}\right)-
\frac{\lambda(\lambda+3)}{r^2}R+
\frac{2M}{\hbar^2}\left(\epsilon + \frac{e^2}{r}\right)R=0
\end{eqnarray}
with the separation constant $\lambda(\lambda+3)$ equal to the nonnegative
eigenvalues of the hypermomentum operator.

It is convenient to make in Eq.(16) a change of variables,
$y=(1-\cos \theta)/2$ and write
\begin{eqnarray*}
Z(y) = y^L(1-y)^JW(y)
\end{eqnarray*}
Substituting this into Eq.(16), we obtain the hypergeometric equation
\begin{eqnarray*}
y(1-y)\frac{d^2W}{dy^2} + [c-(a+b+1)y]\frac{dW}{dy}-abW=0
\end{eqnarray*}
with $a=-\lambda+L+J$, $b=\lambda+L+J+3$, $c=2L+2$.

Thus, we find that
\begin{eqnarray*}
Z(\theta)=(1-\cos\theta)^L(1+\cos\theta)^J \nonumber \\ [2mm]
{_2F}_1\left(-\lambda+J+L,\lambda+J+L+3;2L+2;\frac{1-\cos\theta}
{2}\right)
\end{eqnarray*}
This solution is well behaved at $\theta=\pi$ if the series ${_2F}_1$
terminates, i.e.
\begin{eqnarray*}
-\lambda+J+L = -n_{\theta}
\end{eqnarray*}
where $n_{\theta}=0,1,2,....$

\vspace{1cm}
{\large{\bf 7. Energy Levels.}}
\vspace{0.3cm}

Let us now turn to the radial equation and introduce the function
\begin{eqnarray*}
f(r)=e^{\kappa r}r^{-\lambda} R(r)
\end{eqnarray*}
It is easy to verify that the equation for $f(r)$ has the form of
the confluent hypergeometric equation
\begin{eqnarray*}
z\frac{d^2f}{dz^2} + (c-z)\frac{df}{dz}-af=0
\end{eqnarray*}
where $z=2\kappa r$, $\kappa=\sqrt{-2M\epsilon/\hbar^2}$,
$c=2\lambda+4$, $a=\lambda+2-1/{\kappa r_0}$ and $r_0=\hbar^2/Me^2$.
For the bound state solutions $(\epsilon < 0)$
\begin{eqnarray*}
\lambda+2-1/{\kappa r_0} = -n_r = 0,-1,-2,...
\end{eqnarray*}
and therefore
\begin{eqnarray*}
{\epsilon}_N^T=-\frac{Me^4}{2\hbar^2(\frac{N}{2}+2)^2}
\end{eqnarray*}
where
\begin{eqnarray*}
N=2(n_r+\lambda)=2(n_r+n_{\theta}+J+L)
\end{eqnarray*}

\vspace{1cm}
{\large{\bf 8. Degeneracy.}}
\vspace{0.3cm}

For fixed $T$, the energy levels ${\epsilon}_N^T$ do not depend
on $L$, $J$ and $\lambda$, i.e. are degenerate.
The total degeneracy is
\begin{eqnarray*}
g_N^T =(2T+1)\sum_{\lambda}\sum_{L}(2L+1)
\sum_{J}(2J+1)
\end{eqnarray*}
Since $\lambda=n_{\theta}+J+L$ and $N=2(n_r+\lambda)$, it follows that
(for fixed $N$ and $T$) $\lambda=T,T+1,...,N/2$. Then,
$L_{max}=\lambda-J_{min}$ ($L_{max}$ is fixed) and therefore
$L_{max}=\lambda-(L_{max}-T)$ or $L_{max}=(\lambda+T)/2$. Thus,
\begin{eqnarray*}
g_N^T =(2T+1)\sum_{\lambda=T}^{N/2}\sum_{L=0,1/2}^{\frac{\lambda-T}{2}}
(2L+1)\sum_{J}(2J+1)
\end{eqnarray*}
Now, comparing $|L-T|\leq J\leq L+T$ and $J\leq \lambda-L$ we
conclude that

\vspace{0.3cm}
(a) $J=|L-T|,|L-T|+1,...,L+T$, for $L=0,\frac{1}{2},...,\frac{\lambda-T}{2}$

\vspace{0.3cm}
(b) $J=|L-T|,|L-T|+1,...,\lambda-L$, for
$L=\frac{\lambda-T+1}{2},...,\frac{\lambda+T}{2}$
\vspace{0.3cm}

\noindent
and rewrite the formula for $g_N^T$ in a more explicit form
\begin{eqnarray*}
g_N^T =(2T+1)\sum_{\lambda=T}^{N/2}\Bigl\{
\sum_{L=0,1/2}^{\frac{\lambda-T}{2}}(2L+1)
\sum_{J=|L-T|}^{L+T}(2J+1) +            \\
\sum_{L=\frac{\lambda-T+1}{2}}^{\frac{\lambda+T}{2}}(2L+1)
\sum_{J=|L-T|}^{\lambda-L}(2J+1)\Bigr\}
\end{eqnarray*}
Finally, after some tedious calculations we obtain the following result:
\begin{eqnarray*}
g_N^T =\frac{1}{12}(2T+1)^2\left(\frac{N}{2}-T+1\right)
\left(\frac{N}{2}-T+2\right) \\ [3mm]
\left\{\left(\frac{N}{2}-T+2\right)
\left(\frac{N}{2}-T+3\right)+2T(N+5)\right\}
\end{eqnarray*}
For $T=0$ and $N=2n$ (even) the r.h.s. of the last formula is equal
to $(n+1)(n+2)^2(n+3)/12$, i.e. to the degeneracy of pure Coulomb
levels. Further, $T=0,1,...N/2$ and $T=1/2,3/2,..N/2$ for even and
odd $N$, respectively. Therefore,
\begin{eqnarray*}
g_N =\sum_{T=0,1/2}^{N/2}g_N^T=\frac{(N+7)!}{7!N!}
\end{eqnarray*}
i.e. we obtain the degeneracy of the energy levels for the 8D isotropic
quantum oscillator.

\vspace{1cm}
{\large{\bf 9. Conclusions.}}
\vspace{0.3cm}

Formulae (5) and (6) together with the ansatz (3) form the duality
transformation mapping of the 8D quantum oscillator into the charge--dyon
system with the $SU(2)$--monopole. Let us stress the meaning we use
for the term duality. Both Eq.(4) and Eq.(9) contain two quantities,
$\omega$ and $E$. For Eq.(4) $\omega$ is the fixed parameter
(coupling constant) and $E$ is the quantity to be quantized
(energy of the 8D oscillator). On the contrary, as it easy to see
from (3), for Eq.(9) $E$ is a fixed parameter (Coulomb coupling constant)
and $\omega$ is the quantity to be quantized (${\omega}^2$--energy of the
final system). Thus, the 8D quantum oscillator and the charge--dyon bound
system with the $SU(2)$--monopole are not identical, but dual to each other.

This type duality is valid not only for the 8D, 4D and 2D oscillators,
but also for oscillator--like systems with the potentials
\begin{eqnarray*}
V(u^2)=c_0+c_1u^2+W(u^2)
\end{eqnarray*}
where $W(u^2)$ has a polynomial form
\begin{eqnarray*}
W(u^2)=\sum_{n=2}^{\infty}c_nu^{2n}
\end{eqnarray*}
For such modified potentials, the ansatz (3) can be rewritten as
\begin{eqnarray*}
\epsilon=-\frac{c_1}{4},\,\,\,\,\,\,\,e^2=\frac{E-c_0}{4}
\end{eqnarray*}
Thus, the value of the function $V(u^2)$ at $u^2=0$ contributes to the
Coulomb coupling constant $e^2$. It is also easy to verify that
the l.h.s. of Eq.(9) acquires the additional term ($-W(r)/4r$).

\vspace{1cm}
{\large{\bf 9. Appendix.}}
\vspace{0.3cm}

Consider the normalization of the wave function
$\psi(\vec x,\alpha_T,\beta_T,\gamma_T)$. A standard calculation
shows that the radial wave function $R(r)$ normalized by the condition
\begin{eqnarray*}
\int_0^\infty r^4\left[R_{n_r\lambda}(r)\right]dr=1
\end{eqnarray*}
has the form
\begin{eqnarray}
R_{n_r\lambda}(r)=\frac{4}{r_0^{5/2}(n_r+\lambda+2)^6}
\frac{1}{(2\lambda+3)!}
\sqrt{\frac{(n_r+2\lambda+3)!}{(n_r)!}} \nonumber \\ [2mm]
(2\kappa r)^\lambda e^{-\kappa r}
F\left(-n_r; 2\lambda+4; 2\kappa r\right)
\end{eqnarray}
The full wave function
\begin{eqnarray*}
\psi=C_{LTJ}^{\lambda}R_{n_r \lambda}(r)Z_{\lambda LJ}(\theta)
G_{LTm't'}^{JM}(\alpha,\beta,\gamma;\alpha_T,\beta_T,\gamma_T)
\end{eqnarray*}
is normalized by the condition
\begin{eqnarray*}
\int|\psi|^2dv=1
\end{eqnarray*}
where
\begin{eqnarray*}
dv=r^4\sin^3\theta dr d\theta d\Omega d\Omega_T
\end{eqnarray*}
and
\begin{eqnarray*}
d\Omega=\frac{1}{8}\sin \beta d\beta d\alpha d\gamma,\,\,\,\
d\Omega_T=\frac{1}{8}\sin\beta_Td\beta_Td\alpha_Td\gamma_T
\end{eqnarray*}
Using the formula
\begin{eqnarray*}
{_2F}_1\left(-n,n+a+b+1;a+1;\frac{1-y}{2}\right)
=\frac{n!\Gamma(a+1)}{\Gamma(n+a+1)}P_n^{(a,b)}(y)
\end{eqnarray*}
where $P_n^{(a,b)}(y)$ are the Jacobi polynomials, and taking into
account that
\begin{eqnarray*}
\int_{-1}^1(1-y)^a(1+y)^b\{P_n^{(a,b)}(y)\}^2dy= \nonumber \\ [2mm]
\frac{2^{a+b+1}}{2n+a+b+1}\frac{\Gamma(n+a+1)\Gamma(n+b+1)}
{n!\Gamma(n+a+b+1)}
\end{eqnarray*}
\begin{eqnarray*}
\int D_{m_2m'_2}^{j_2^*}(\alpha,\beta,\gamma)
D_{m_1m'_1}^{j_1}(\alpha,\beta,\gamma)d\Omega=
\frac{2\pi^2}{2j_1+1}
\delta_{j_1j_2}\delta_{m_1m_2}\delta_{m'_1m'_2}
\end{eqnarray*}
it is easy to obtain that
\begin{eqnarray*}
C_{LJT}^\lambda = \sqrt{\frac
{(2L+1)(2T+1)(2\lambda+3)(\lambda-J-L)!\Gamma(\lambda+J+L+3)}
{2^{2J+2L+5}\pi^4\Gamma(\lambda+J-L+2)\Gamma(\lambda-J+L+2)}}
\end{eqnarray*}

\vspace{1cm}
{\large{\bf 11. Acknowledgement.}}
\vspace{0.3cm}

We are grateful to Yeranuhy Hakobyan, Armen Nersessian and George
Pogosyan for many discussions on the subject of dyon--oscillator
duality.


\begin{thebibliography}{99}
\bibitem{1}
N.Seiberg, E.Witten. Nucl.Phys. B, 1994, vol. 431, p. 484.
\bibitem{2}
L.G.Mardoyan, A.N.Sissakian, V.M.Ter--Antonyan. \\
Oscillator as a Hidden Non--Abelian Monopole. Preprint JINR
E2-96-24, Dubna, 1996. Hep--th/9601093.
\bibitem{3}
V.Ter--Antonyan, A.Nersessian. Mod.Phys.Lett. A, 1995, vol. 10, p. 2623.
\bibitem{4}
A.Nersessian, V.Ter--Antonyan, M.Tsulaya. \\
Mod.Phys.Lett. A, 1996, vol. 11, p. 1605.
\bibitem{5}
A.Magkakian, A.N.Sissakian, V.M.Ter--Antonyan. \\
Phys.Lett. A, 1997, in press.
\bibitem{6}
T.Iwai, T.Sunako. J.Geom.Phys., 1996, vol. 20, p. 250.
\bibitem{7}
L.S.Davtyan, L.G.Mardoyan, G.S.Pogosyan, A.N.Sissakian, \\
V.M.Ter--Antonyan. J.Phys. A, 1987, vol. 20, p. 6121.
\bibitem{8}
P.Kustaanheimo, E.Stiefel. J.Reine Angew. Math., 1965, vol. 218, p. 204.
\bibitem{9}
D.Zwanziger. Phys.Rev., 1968, vol. 176, p. 1480.
\bibitem{10}
D.Lambert, M.Kibler. J.Phys. A, 1988, vol. 21, p. 307.
\bibitem{11}
Le Van Hoang, Tony J.Viloria, Le anh Thu.
J.Phys. A, 1991, vol. 24, p. 3021.
\bibitem{12}
C.N.Yang. J.Math.Phys., 1978, vol. 19, p. 320.

\end{thebibliography}
\end{document}